\documentclass[aip, pop, amsmath,amssymb, reprint,floatfix]{revtex4-1}

\usepackage{color}
\usepackage{graphicx}
\usepackage{dcolumn}
\usepackage{bm}
\usepackage{psfrag}
\usepackage{epstopdf}
\usepackage{subfigure}
\usepackage{textcomp}
\usepackage{hyperref}
\usepackage{comment}
\usepackage{multirow}
\usepackage{amstext} 
\usepackage{amsmath}

\usepackage{sidecap}

\newcommand{\Baxis}{$B_{axis}$}
\newcommand{\Bperp}{$B_{\perp}$}
\newcommand{\sbend}{$\epsilon_{bend}$}
\newcommand{\stot}{$\epsilon_{tot}$}
\newcommand{\stor}{$\epsilon_{tor}$}
\newcommand{\snot}{$\epsilon_{0}$}
\newcommand{\thwind}{$\theta_{wind}$}
\newcommand{\ravg}{$\left<r_{coil}\right>$} 
\newcommand{\size}{size scale factor}
\newcommand{\Icrit}{$I_{crit}$}
\newcommand{\Ltape}{$L_{tape}$}
\newcommand{\dg}{$^{\circ{}}$}
\newcommand{\easy}{major-axis}
\newcommand{\wrong}{minor-axis}
\newcommand{\uHTS}{NI-HTS}

\newcommand{\red}{{\color{red}}}

\begin{document}

\title[]{Non-Planar Coil Winding Angle Optimization for Compatibility with Non-Insulated High-Temperature Superconducting Magnets}
\author{C. Paz-Soldan}
\affiliation{General Atomics, San Diego, California 92186-5608, USA}
\email{paz-soldan@fusion.gat.com}



\date{\today}

\begin{abstract}
The rapidly emerging technology of high-temperature superconductors (HTS) opens new opportunities for the development of non-planar non-insulated HTS magnets. This type of HTS magnet offers attractive features via its simplicity, robustness, and is well-suited for modest size steady-state applications such as a mid-scale stellarator.  In non-planar coil applications the HTS tape may be subject to severe \wrong{} bending strain (\sbend{}), torsional strains (\stor{}) and transverse magnetic field components (\Bperp{}), all of which can limit the magnet operating space. A novel method of winding angle optimization is here presented to overcome these limitations. Essentially, this method: 1) calculates the peak \sbend{} and \Bperp{} for arbitary winding angle along an input coil filamentary trajectory, 2) defines a cost function including both, and then 3) uses tensioned splines to define a winding angle that reduces \stor{} and optimizes the \sbend{} and \Bperp{} cost function. As strain limits are present even without \Bperp{}, this optimization is able to provide an assessment of the minimimum buildable size of an arbitrary non-planar non-insulating HTS coil. This optimization finds that for standard 4 mm wide HTS tapes the minimum size coils of the existing HSX, NCSX, and W7-X stellarator geometries are around 0.3 - 0.5 m in radius. For coils larger than this size, permitting a finite (yet tolerable) strain allows reduction of \Bperp{}. This enables a reduction of the HTS tape length required to achieve a given design magnetic field or equivalently an increase in the achievable magnetic field for fixed HTS tape length. The distinct considerations for optimizing a stellarator coilset to further ease compatibility with non-insulated HTS magnets are also discussed.
\end{abstract}


\maketitle



\section{Introduction and Motivation}
\label{sec:intro}

High-temperature superconductors (HTS) have been recognized for the past two decades to offer attractive new pathways for magnet development \cite{Bruzzone2018}. Compared to low-temperature superconductors (LTS), HTS enables the design of magnets that operate either at higher magnetic field, higher temperature, higher current density, or combinations of all three. Compared to copper, HTS (and LTS) offers the benefit of significantly reduced energy dissipation within the magnet, enabling continuous operation at higher magnetic field. Naturally, these attributes of HTS are opening new opportunities for applications that benefit from improved magnets \cite{Haught2007,Fietz2013, Whyte2016, Maingi2019}. Robust efforts are ongoing to deploy HTS technology toward large-bore high-field magnets for magnetic fusion energy applications \cite{Sorbom2015,Sykes2018}.

While worldwide focus has been largely directed towards high-field planar magnet systems, little attention has been paid to new opportunities enabled by HTS in applications that benefit from improved non-planar magnets. Non-planar configurations can be found in force-balanced (helical) coils for magnetic energy storage \cite{Miura1994}, particle accelerator magnets using saddle/bedstead \cite{Thomas2005} or canted cosine theta \cite{Amemiya2015} geometries, and the stellarator concept of a magnetic fusion energy system \cite{Najmabadi2006,Wolf2008}.

The unusual (essentially 2D) form factor of HTS tape has given rise to several methods to convert the tape into a viable conductor. Integrated multi-tape conductor concepts include: interleaving HTS tapes into a Roebel assembly \cite{Goldacker2007}, winding HTS tape helically along a cylindrical form (termed Cable on Round Core, CORC conductor \cite{Weiss2017}), and forming stacks of many HTS tape layers and winding the stack in various arrangements (termed twisted stacked-tape conductor \cite{Takayasu2012}). However, the first and still the simplest method to construct a magnet from HTS tape is simply to wind the HTS tape in a `bare' non-insulated and non-epoxy impregnated configuration around a bobbin that defines the shape of the final coil. This type of coil is referred to as a non-insulated HTS coil (\uHTS{})\cite{Hahn2011,Kim2012a}.

\subsection{Primer on Benefits and Drawbacks of Uninsulated HTS Magnets}

A central benefit of the \uHTS{} magnet is its simplicity. In this configuration the HTS tape is wound directly onto a shaped bobbin that defines the winding geometry, and the desired performance (in kiloamp turns, kAt) is achieved simply by adding turns. These coils do not require (and indeed cannot allow) cooling channels within the conductor stack. Any heat generated must instead be rejected through the bobbin structure. Also, as the number of turns in the \uHTS{} magnet is generally very large, a low supply current (and thus a low input power) is required to drive them.

Beyond simplicity, owing to the absence of an insulator between turns, \uHTS{} magnets offer a degree of intrinsic superconductivity quench protection, as the electrical current is offered a multitude of parallel paths to avoid the non-superconducting failure point \cite{Hahn2016}. Finally, as the HTS tape itself consists of superconducting layers deposited onto a steel substrate, winding an \uHTS{} tape magnet on a steel bobbin results in a final assembly mechanically very similar to pure steel. This yields reduced differential thermal expansion issues and significantly enhanced strength as compared to other magnets.

Drawbacks can also be identified. Owing to the large number of turns of conductor ($N$) required, \uHTS{} coils are typically high in inductance ($L \propto N^2$) and thus cannot change current quickly unless large voltages are utilized. This makes \uHTS{} coils challenging to use in AC operation modes. This often limits deployment to truly steady-state applications, such as long-timescale energy storage, particle accelerators, and the stellarator fusion concept. Notably, \uHTS{} would be challenging to use in the poloidal field coils of the tokamak fusion concept due to the time-varying current requirement.

The \uHTS{} coil also suffers from a second drawback. For large-bore, high-field applications, the number of turns required is very large, as is the path length of each turn. Either severely long lengths of HTS tape or a large number of resistive joints are thus required, creating a practical limitation to the ultimate potential of this magnet type. These drawbacks naturally drive development towards the complex multi-tape conductor assemblies as described earlier.

\subsection{Compatibility of Uninsulated HTS Magnets with Non-Planar Applications}

Considering deployment of \uHTS{} magnets to non-planar applications, two additional constraints arise. First, the radius of curvature along the winding trajectory no longer points towards a fixed point, but instead can take arbitrary form. This necessitates the introduction of {\it{wrong-ways bending}} strain and {\it{torsional bending}} strain. Secondly, the magnetic field generated by the magnet is no longer predominantly parallel to the HTS tape plane (as it is in a planar magnet), but instead has significant transverse field components (\Bperp{}). Both of these issues degrade the HTS tape performance and ultimately limit its operating space.

In this work a new winding angle optimization method is developed and presented to mitigate the aforementioned HTS tape compatibility issues of strain and transverse field. The winding angle is a free parameter for any filamentary coil model, and will here be exploited as an optimization parameter to mitigate the issues associated with deploying \uHTS{} coils in non-planar applications.

\subsection{Goal, Structure, and Summary of Work}

The goal of this paper is to discuss the compatibility of \uHTS{} coils for non-planar applications and to present a novel winding angle optimization method developed to overcome the identified limitations. The optimization method is described in Sec. \ref{sec:method}, and the candidate non-planar coil geometries examined (well-known stellarator designs) are described in Sec. \ref{sec:cases}. Results of strain-only optimizations are presented in Sec. \ref{sec:strain}. These optimizations are able to assess the minimum size of a non-planar coil that can be wound without exceeding strain limits for a given width of HTS tape, which are found to be 0.3 - 0.5 m radius for the studied stellarator configurations. Results of combined strain and \Bperp{} optimizations are presented in Sec. \ref{sec:both}. By defining coils larger than the minimum size, headroom is created to allow reduction of the \Bperp{} component, enabling access to higher field for fixed tape HTS length, or the same field at reduced HTS tape length. The degree of benefit depends on the target coil size, as this method can quantify. Conclusions are presented in Sec. \ref{sec:disc} along with a discussion of how to optimize the coil geometry itself for improved compatibility with \uHTS{} magnets.


\section{Winding Angle Optimization Method}
\label{sec:method}

By calculating the peak strain due to \wrong{} bending (\sbend{}) along with \Bperp{} along the coil trajectory as a function of winding angle (\thwind{}), a trajectory can be found that minimizes arbitrary cost functions of these two metrics. To minimize torsional strain (\stor{}) a tensioned spline fit to the optimal trajectory allows identification of the optimum trade-off between the cost function and \stor{}. Each of these steps is now described in detail.

\subsection{Strain Considerations}
\label{sec:strlimit}
\begin{figure*}
\centering
\includegraphics[width=0.89\textwidth]{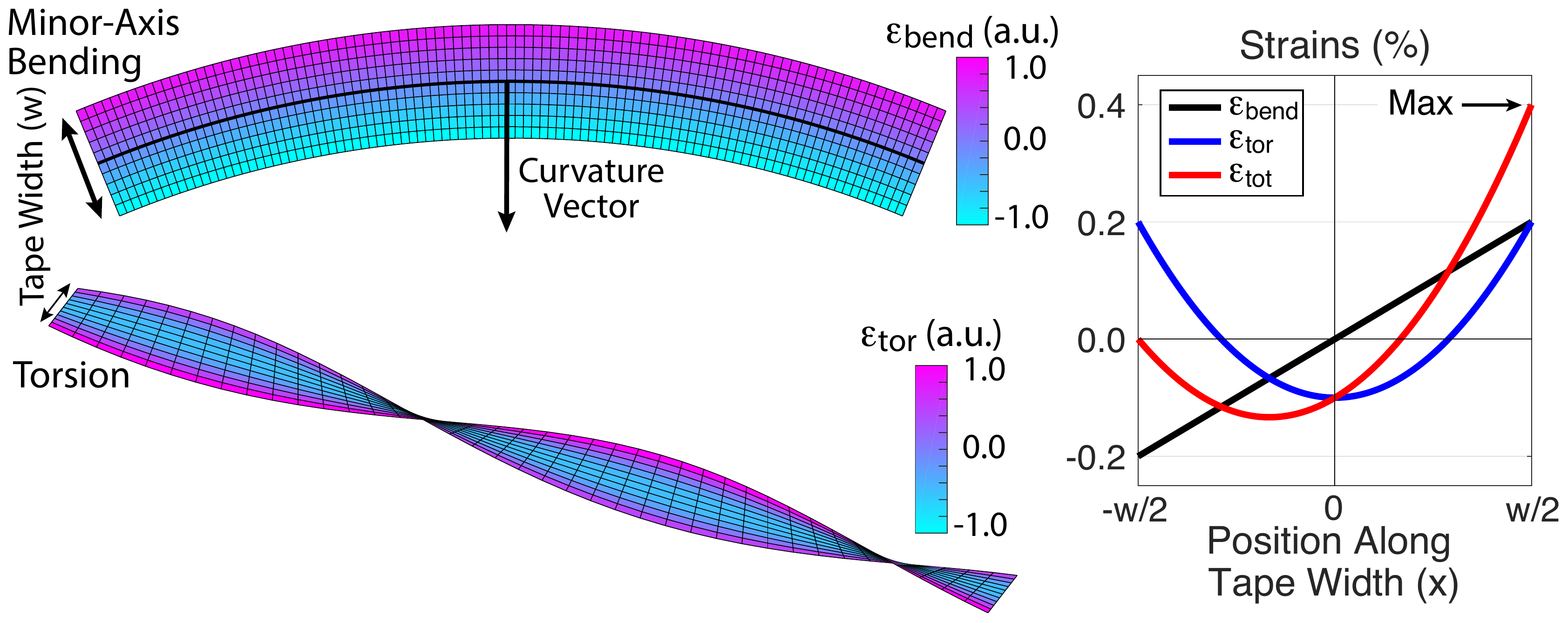}
\caption{Visualization of the strain components considered in the optimization. The \wrong{} bending strain (\sbend{}, top-left) is linearly proportional to the distance along the HTS tape width, while the torsional strain (\stor{}, bottom-left) takes a offset-parabolic form. The peak strain (\stot{}) is simply found by summing these two components, and it is always found at one edge of the HTS tape.}
\label{fig:basic}
\end{figure*}

Two strain components are possibly severe in non-planar coils made with \uHTS{} tape, as illustrated in Fig. \ref{fig:basic}. These are the \wrong{} bending strain (\sbend{}) and the torsional strain (\stor{})\cite{Takayasu2010}.

The \wrong{} bending strain is linearly proportional to the distance along the HTS tape width, and the magnitude depends on the radius of curvature via the following simple relationship:

\begin{equation}
\epsilon_{bend}(x) = \frac{x}{\left| r_{C} \right|},
\label{eq:sbend}
\end{equation}
where $r_{C}$ is the radius of curvature, and $x$ is the position along the tape width $w$. It peaks at $x=w/2$, the tape edge, with a value of \sbend$=w/2|r_{C}|$. Here $r_{C}$ is calculated numerically using finite differences \cite{Wang2017a}, though if the coil trajectory is parametrized it can also be described analytically using the Frenet-Serret formulas \cite{Gray2006}.

Note the \easy{} bending strain is also given by a similar relation, but it is smaller by the ratio of the tape width to its thickness. As normal HTS tape widths are 4, 6, and 12 mm while thicknesses are 0.1 mm, this strain component can be safely ignored. This also means that as long as the radius of curvature is directed along the major axis of the tape, a 40x-120x smaller radius of curvature can be tolerated. This can greatly impact optimization of the coil trajectory itself as will be discussed separately in Appendix \ref{sec:stellopt}. Note \easy{} strain is ignored in this study because it is so much lower than \wrong{} strain.

The torsional strain (\stor{}) does not depend on the local radius of curvature but instead is related to the angular rate of change of $r_{C}$ along the coil trajectory. The torsional strain takes the form \cite{Takayasu2010}:
\begin{equation}
\epsilon_{tor}(x) = \frac{1}{2}\left(\frac{\Delta \theta_{wind}}{\Delta L}\right)^2 \left(x^2-\frac{w^2}{12}\right)
\label{eq:stor}
\end{equation}
for a position $x$ along the tape width $w$, where $\Delta \theta$ is the angular rate of change of the winding angle \thwind{} per unit length along the coil trajectory ($\Delta L$). \stor{} takes an offset-parabolic form and also peaks at the tape edge ($x=\pm w/2$), with a value of $\epsilon_{tor}=(\Delta \theta_{wind} / \Delta L)^2 w^2/12$. $\Delta \theta_{wind}$ is also calculated numerically using finite differences, but it too can be described analytically using the Frenet-Serret formulas if the trajectory is parametrized.

A scalar metric representing the total strain ($\equiv$\stot{}) is now defined from \sbend{} and \stor{}. To rigorously treat the problem, the three-dimensional internal strains at every point in the tape should be taken into account using the Principle Strain Method (PSM) \cite{Young2011}, including actual material properties such as the Poission Ratio, Modulus of Elasticity, and Modulus of Rigidity to relate the different strain tensor elements. To simplify the problem and avoid sensitivity to material properties, a less rigorous but more conservative metric is used in this work - the maximum of a scalar sum of the \sbend{} and \stor{} components:
\begin{equation}
\epsilon_{tot}=\max( \epsilon_{bend}(x)+\epsilon_{tor}(x)).
\label{eq:stot}
\end{equation}
Note the maximum \stot{{} always occurs at one tape edge or another ($x=\pm w/2$), based on the relative directions of \sbend{} and \stor{}. This method is conservative because it assumes the strain components are fully co-linear (which is approximately true in the limit of thin tapes). Comparison of Eq. \ref{eq:stot} and the 	PSM finds the strain can be over-estimated by 15-20\% by Eq. \ref{eq:stot}. This over-estimate is expected to be compensated by increases in the real material strain introduced by material imperfections, giving additional credence to this conservative approach. 

In terms of a limit to the acceptable \stot{}, in principle empirical data should be gathered at the target operating strain and field conditions to validate the expected performance of the HTS tape. In the absence of such data this study uses an industry rule-of-thumb, which is that a maximum \stot{} limit of 0.4\% should be enforced \cite{Takayasu2015,Allen2015}. Above this limit there is a risk of reduction in the critical superconducting current \Icrit{} capacity as well as delamination of the internal layers within the HTS tape \cite{Zhang2016}. Regardless, the optimization framework can take arbitrary strain limits as input, and results are generally given in terms of peak predicted \stot{}.

\subsection{Transverse Field Considerations}
\label{sec:bperplimit}
While strain is the primary consideration as exceeding its limit has more severe consequences, a secondary consideration is the transverse magnetic field (\Bperp{}). This imposes a soft limit on HTS tape performance as increasing \Bperp{} also degrades \Icrit{}. For this study, data on this limitation is obtained from publicly available HTS tape manufacturer data \cite{Superpower}.

Note that unlike the strains, \Bperp{} depends on coils throughout the entire configuration. As such to compute \Bperp{} the fields from all conductors in the configuration must be taken into account. This includes all other magnets as well as the fields from other turns within the magnet. A limitation of the present study is that the cross-sectional geometry of the conductor is not defined, as such the \Bperp{} from finite coil winding pack aspect ratio is not presently treated. This limitation may be improved upon in the future.

\subsection{Optimization Philosophy and Cost Function Definition}
\label{sec:cost}

\begin{figure}
\centering
\includegraphics[width=0.43\textwidth]{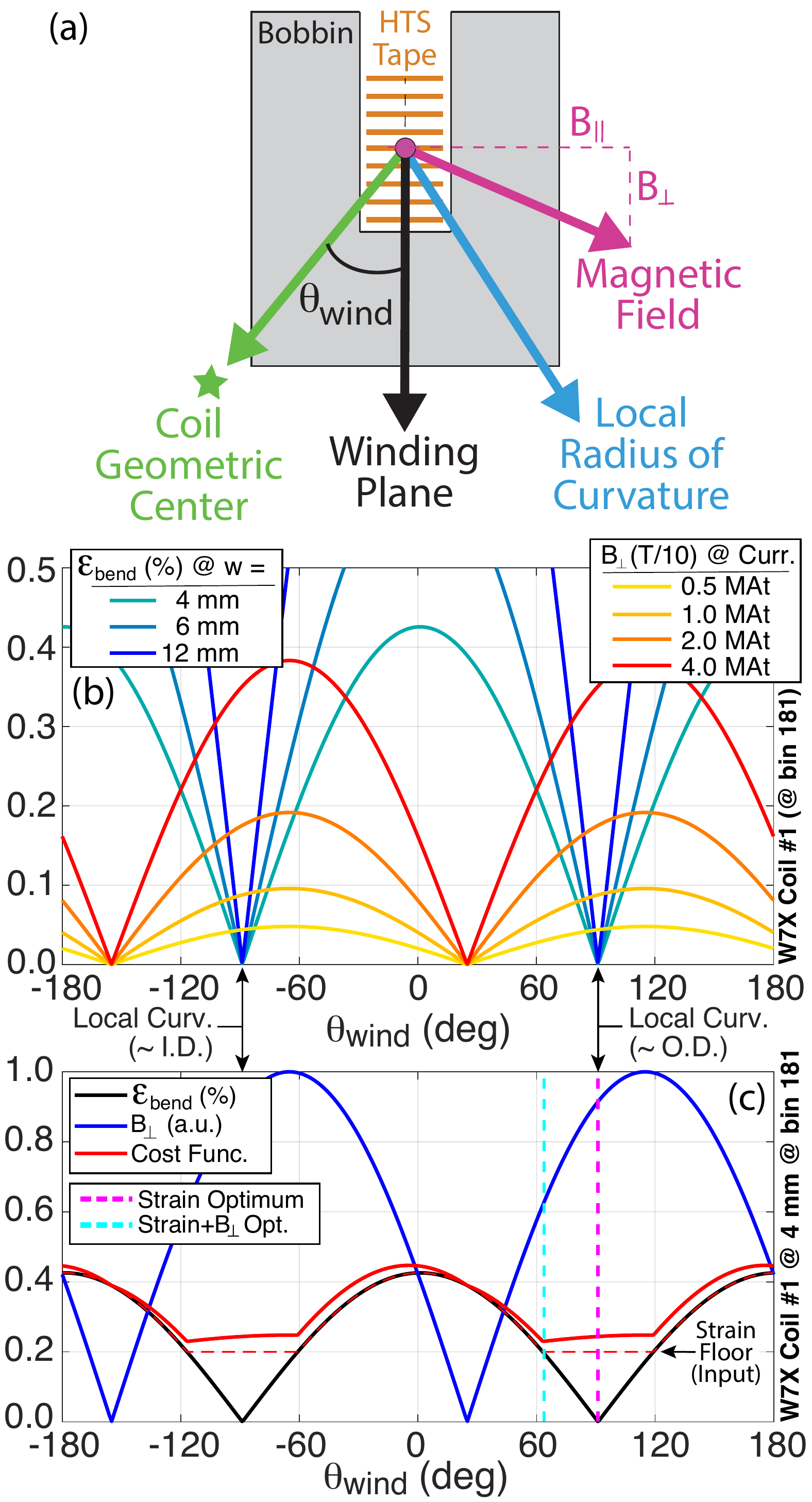}
\caption{(a) Cartoon illustration of a \uHTS{} coil and example orientations of the winding angle (\thwind{}), the magnetic field direction, and the local radius of curvature. Note \thwind{} is defined relative to the coil geometric center. (b) Example evaluations of the \wrong{} bending strain (\sbend{}) and transverse field (\Bperp{}) as a function of \thwind{}, with minima of each occuring for different \thwind{}. \sbend{} depends on HTS tape width while \Bperp{} depends on the coil current. If \thwind{} is aligned to the local curvature then \sbend{}=0. (c) Construction of a cost function (Eq. \ref{eq:cost}) allowing \Bperp{} to be reduced while maintaining \sbend{} below an input tolerable strain floor \snot{}. Solutions exist on both the bobbin effective outer diameter (O.D) and inner diameter (I.D.)}
\label{fig:1Dplot}
\end{figure}

As the \sbend{} limit is a hard constraint on the HTS integrity, while \Bperp{} is a softer limit, the optimization philosophy is thus to first ensure strain is within tolerable limits, and then within these limits to optimize against \Bperp{} as a secondary constraint.  Since \sbend{} and \Bperp{} are single-valued functions of the winding angle \thwind{}, they can be directly computed for all possible \thwind{}. This is shown in Fig. \ref{fig:1Dplot}(b) for a single point along an example coil trajectory (the coil geometries considered will be described in Sec. \ref{sec:cases}). As can be seen, \sbend{} depends sensitively on HTS tape width, while \Bperp{} naturally depends on the coil current. As can also be seen, the optimal \thwind{} to minimize \Bperp{} and \sbend{} are different. Aligning \thwind{} to the local curvature ensures \sbend{}=0, noting this can be achieved on either the bobbin effective outer diameter (O.D.) or inner diameter (I.D.).

The method chosen to enable simultaneous optimization of \sbend{} and \Bperp{} is to define a cost function that is a linear sum of \sbend{} and \Bperp{} with an ad-hoc relative scale-factor $\alpha$:
\begin{equation}
Cost =
\begin{cases}
  \epsilon_{bend} + \alpha B_{\perp} & \text{if} \ \epsilon_{bend}> \epsilon_{0} \\\\
  \epsilon_{0} + \alpha B_{\perp} & \text{if} \ \epsilon_{bend}< \epsilon_{0}   \\
  \end{cases}
\label{eq:cost}
\end{equation}
The parameter \snot{} is the bending strain that is deemed to be tolerable and is an input free parameter.  An essential feature of the cost function is that when \sbend{} is below \snot{} the cost function sees no variation arising from \sbend{}. An example cost function as applied to {\red{the same}} coil configuration is shown in Fig. \ref{fig:1Dplot}(c). As can be seen, as long as the relative scale factor $\alpha \ll max(B_{\perp})/max(\epsilon_{bend})$, the \Bperp{} term will only have an effect when \sbend{} $<$ \snot{}, as desired. If \Bperp{} considerations are ignorable, setting $\alpha=0$ results in a cost function equal to only \sbend{}.

\subsection{Torsion Optimization via Tensioned Splines}
\label{sec:spline}

Definition of a cost function to optimize \thwind{} is not sufficient to solve the optimization problem, as the actual \thwind{} trajectory itself impacts the total strain via the torsional strain \stor{}. This is because \stor{} is related to the rate of change of \thwind{} along the trajectory. To address this problem, the approach is to compute the cost function across all possible \thwind{} at all positions along the coil trajectory. This gives rise to contour plots of the cost function that visualizes the optimization problem, and provides a graphical method to reduce \stor{} while minimally increasing the cost function.

\begin{figure}
\centering
\includegraphics[width=0.45\textwidth]{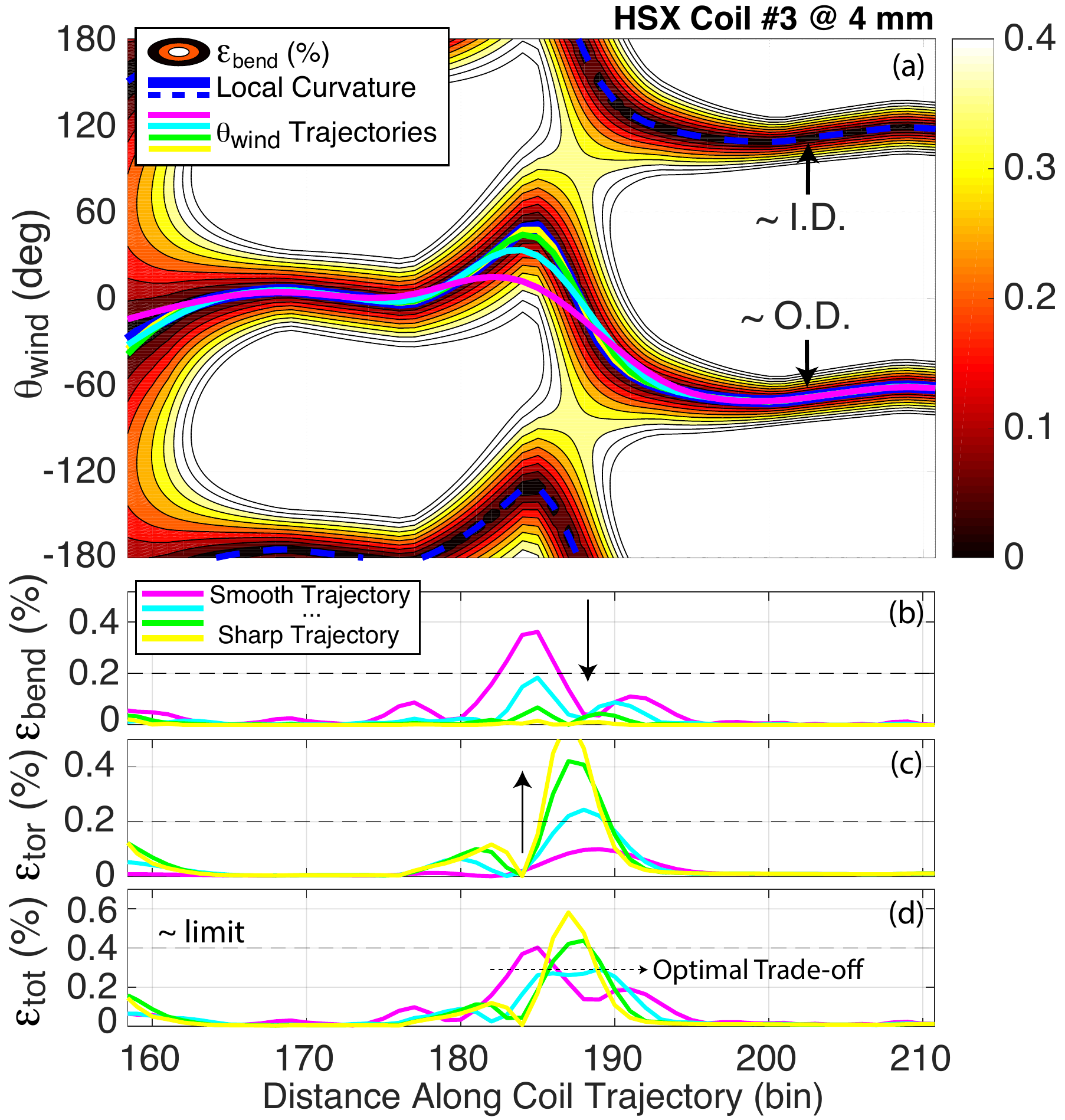}
\caption{Example use of spline tension to minimize total strain. (a) The cost function (here \sbend{}) is plotted for all winding angles (\thwind{}) for a subset of an example coil trajectory. Varying the spline tension yields various possible \thwind{} trajectories. (b-d) These different trajectories trade off \sbend{} and \stor{} differently, giving rise to an optimum in the total strain (\stot{}).}
\label{fig:tension1}
\end{figure}

To simply illustrate this step of the optimization process a \sbend{}-only cost function ($\alpha=0$ in Eq. \ref{eq:cost}) is used, and only a subset of the coil trajectory is shown in Fig. \ref{fig:tension1}. As can be seen, the contours in Fig. \ref{fig:tension1}(a) are simply \sbend{} contours along the coil trajectory for all possible \thwind{}. The final \thwind{} is fit to the minimum of the cost function (the minimum of \sbend{} in Fig. \ref{fig:tension1} using a tensioned spline approach. The magnitude of the local radius of curvature is used as a fitting weight for the tensioned spline, with low curvature regions ascribed a low weight. Additionally, manual adjustment of the fit is possible by inserting points with high weighting to the fitting. This can drive the fit to find alternate optimal paths through the winding trajectory. Different fitted trajectories of \thwind{} are indicated as the colored lines in Fig. \ref{fig:tension1}, with different tensions for each. For low spline tension, the fit closely matches the cost function minimum, while for high tension the variation of \thwind{} along the coil trajectory is minimized. As can be seen in Fig. \ref{fig:tension1}(b)-(d), this allows a direct tradeoff between \sbend{} and \stor{}, and enables a minimum \stot{} (=\sbend{}+\stor{}) to be identified.  Note in some instances the optimal \thwind{} trajectory includes regions where winding is primarily on the inner diameter of the coil (\thwind{} $\approx 180$\dg{}), as opposed to the outer diameter (\thwind{} $\approx 0$\dg{}).

While this method is surely not a unique solution to the optimization problem, the simple treatment is found to be sufficiently flexible to achieve the desired reduction in \Bperp{} within allowable \sbend{} constraints. 

At this point a key difference between this method and the method of calculating space-preserving maps \cite{Gray2006}, giving rise to developable surfaces (also called the constant-perimeter method) should be clarified. As a result of the tensioned spline method utilized here, the optimal winding angle \thwind{} does not necessarily follow the radius of curvature. As such the final tape surface is {\it{not an area-preserving map}}, and indeed this is why finite \sbend{} is present. Were an area-preserving map method map utilized, the resultant trajectory would likely undergo severe \stor{} as a result of its inability to trade off \stor{} with \sbend{}, as is done here. Also to be noted is that the optimization workflow also can treat a planar coil, in which case the optimal \thwind{} returns zero throughout as expected.


\section{Stellarator Coil Configurations Considered}
\label{sec:cases}

\begin{figure}
\centering
\includegraphics[width=0.48\textwidth]{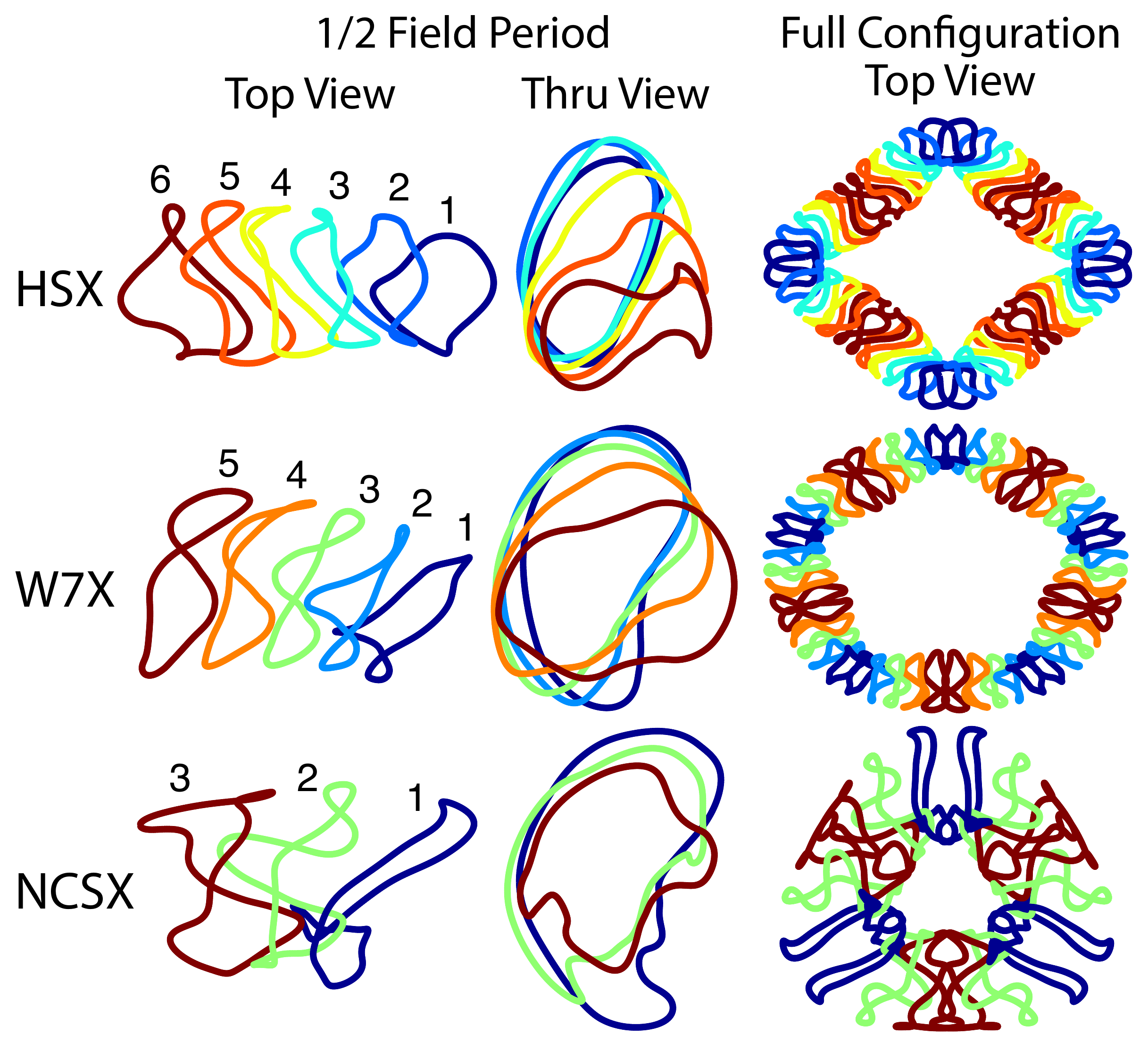}
\caption{Stellarator geometries considered in this study. Non-planar coils in these configurations span from weakly to strongly non-planar.}
\label{fig:stellgeom}
\end{figure}

Though the optimization methods described in Sec. \ref{sec:method} are applicable to arbitrary coil geometry, well-known yet complex coil geometries from the stellarator are used as examples. The configurations studied are the Helically Symmetric Experiment (HSX) \cite{Anderson1995}, the Wendelstein 7-X (W7-X) stellarator \cite{Beidler1990,Klinger2013}, and the National Compact Stellarator Experiment (NCSX) \cite{Chrzanowski2007,Zarnstorff2001}. Each coilset was generated primarily based on varying constraints arising from plasma physics, alongside engineering constraints from the coilsets. Note that each configuration differs in physical size and magnet technology (HSX and NCSX are copper while W7-X is LTS). These coils, along with identifying coil numbers assigned for the purpose of this study, are shown in Fig. \ref{fig:stellgeom}. Note that Fig. \ref{fig:tension1} used the \#3 coil of the HSX configuration.

\begin{figure}
\centering
\includegraphics[width=0.48\textwidth]{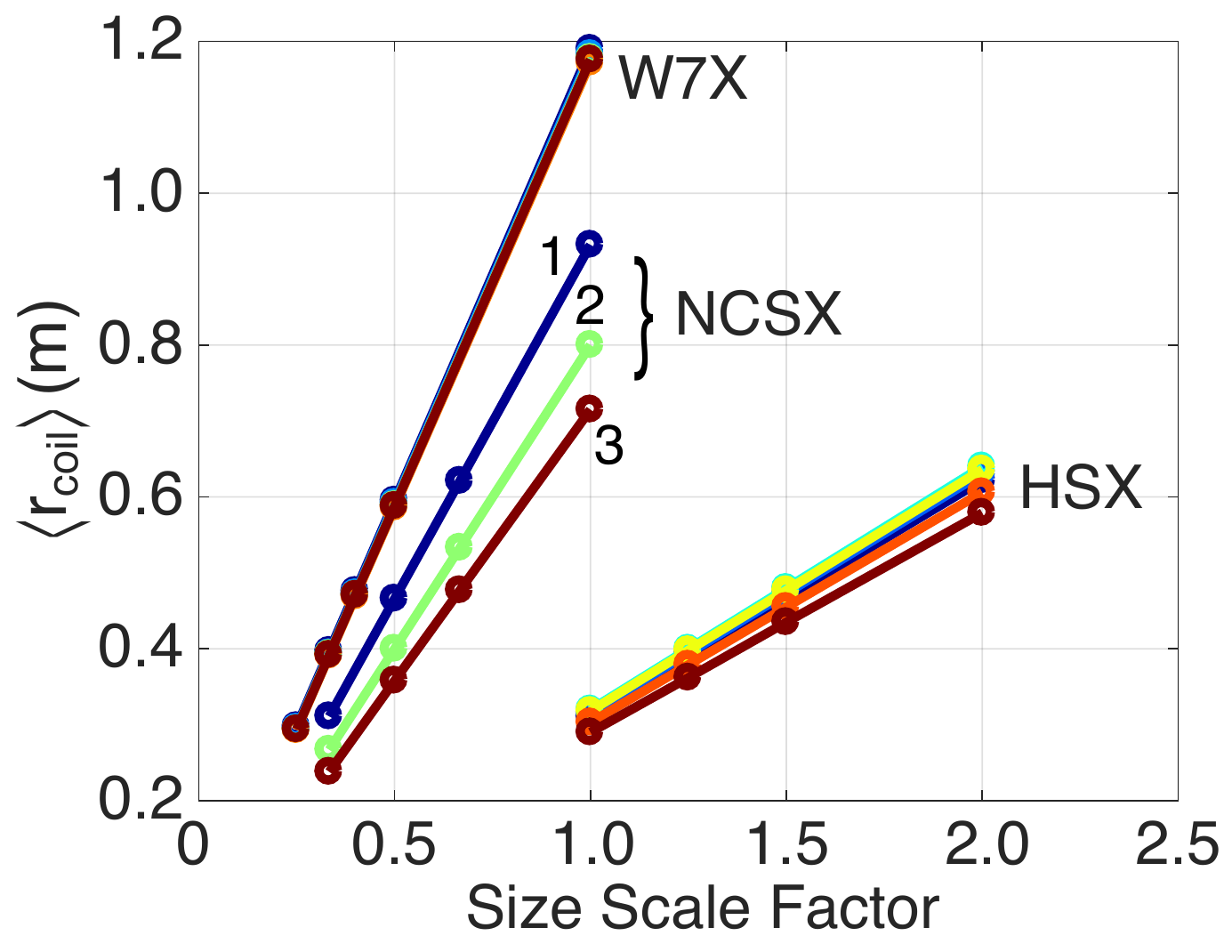}
\caption{Mean coil radius (\ravg{}) for each of the stellarator geometries considered as a function of the \size{} applied. Size scale factor of 1 is the size of the as-built coil.}
\label{fig:stellsize}
\end{figure}

In order to assess sensitivity to coil size and to estimate the minimum buildable coil size, a uniform geometric scale factor was applied to each of the stellarator configurations shown in Fig. \ref{fig:stellgeom}. The average coil radius (\ravg{}) for each of these designs is plotted against the \size{} applied in Fig. \ref{fig:stellsize}.  \ravg{} is defined here as the mean distance from the coil trajectory to the coil geometric center. The coil geometric center is in turn defined as the mean position of the coil trajectory. 

As can be seen, in terms of \ravg{}, W7-X is the largest though NCSX is only modestly smaller. However as can be seen in Fig. \ref{fig:stellgeom}, the complexity of the NCSX coils is considerably increased due to the more stringent constraints utilized in its optimization (in particular the desire for a tight aspect ratio). The HSX coils are smallest and also the most simple. All devices were scaled such that they occupied an overlapping \ravg{} range between 0.2 and 0.6 m.


\section{Strain Optimization and Minimum Coil Size}
\label{sec:strain}

The main objectives of optimizations involving only strain are to provide headroom to further reduce \Bperp{} and to enable the use of progressively wider HTS tape widths (thus increasing the current capacity per turn). Strain-only optimizations also provide a means of determining the minimum buildable size of a \uHTS{} coil at fixed tape width regardless of target \Bperp{}. As described in Sec. \ref{sec:strlimit}, a value of 0.4\% is considered engineering best practice and is here used as the target allowable \stot{}. Results will be conveyed by plotting the peak strain (\stot{}) vs coil size (\ravg{}), in case further HTS advances modify the allowable strain.

\begin{figure}
\centering
\includegraphics[width=0.48\textwidth]{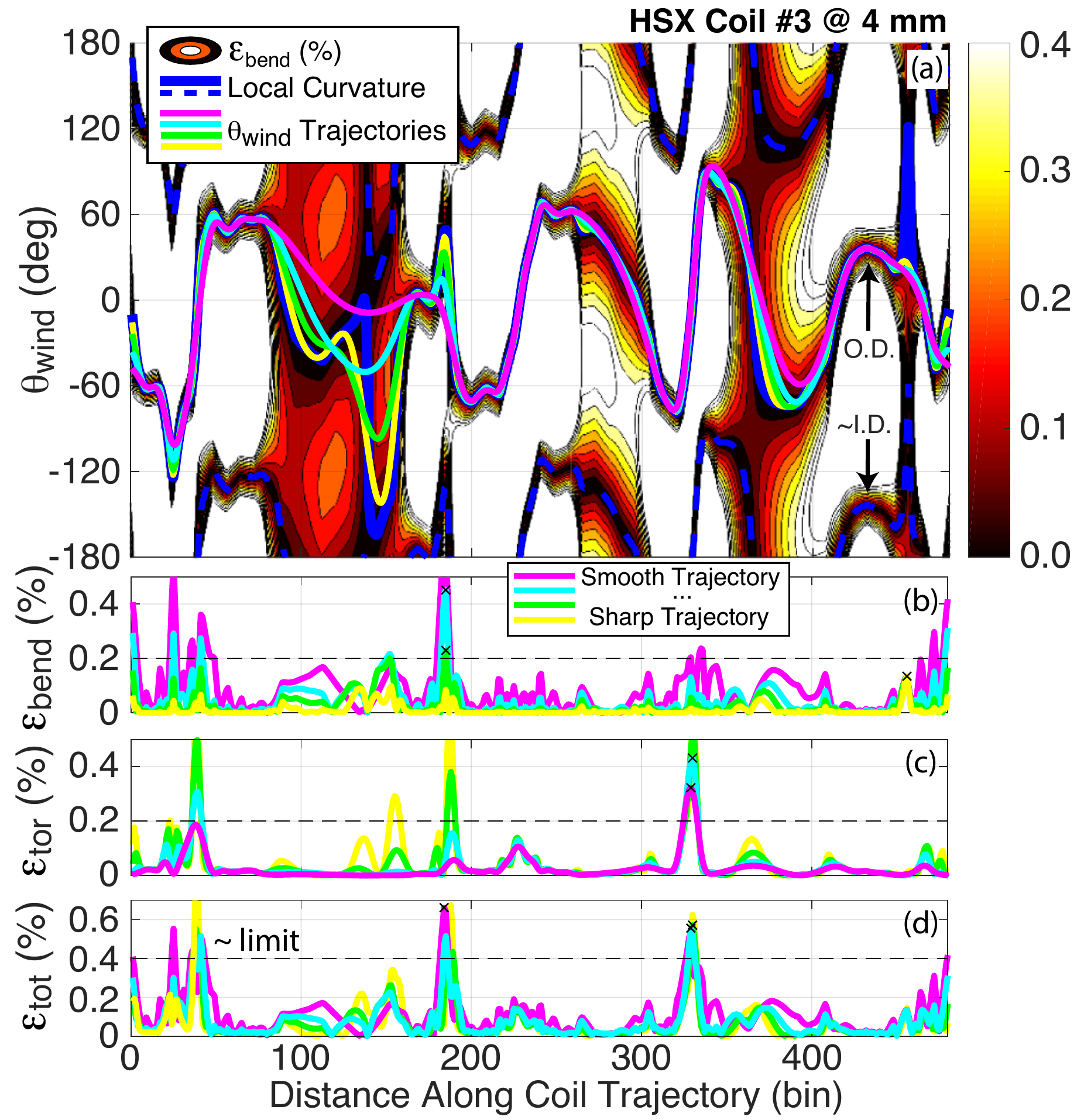}
\caption{The same \thwind{} trajectories of Fig. \ref{fig:tension1} now displayed for the entire HSX \#3 coil trajectory. Again varying the spline tension yields (a) various candidate \thwind{} trajectories. (b-d) These different trajectories trade off \sbend{} and \stor{} differently, giving rise to an optimim in the total strain (\stot{}).}
\label{fig:tension2}
\end{figure}

The full coil trajectory for the \#3 HSX coil shown in Fig. \ref{fig:stellgeom} (and highlighted in Fig. \ref{fig:tension1}) is shown in Fig. \ref{fig:tension2}. For this coil, some regions of the coil trajectory are very strongly constrained by \sbend{} while others are not. The tensioned spline approach allows quick identification of the optimal \thwind{} trajectory.

\begin{figure}
\centering
\includegraphics[width=0.49\textwidth]{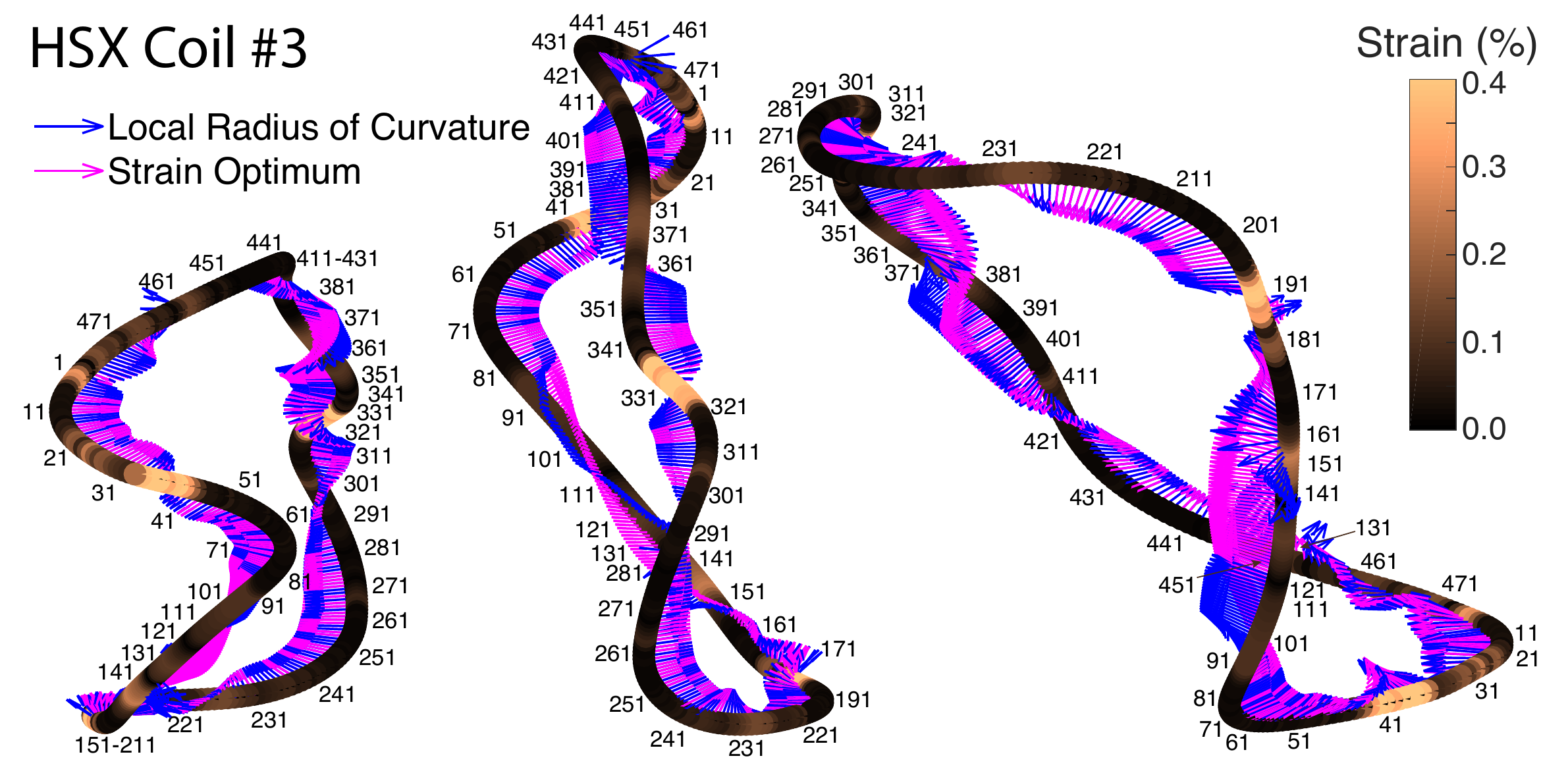}
\caption{Three viewing angles of HSX coil \#3 showing the local radius of curvature (blue vectors) and optimal \thwind{} trajectory (magenta vectors) for a strain-only optimization . Colors along the coil trajectory indicate relative \stot{}. Regions of high \stot{} are found at the transition between bends.}
\label{fig:3dHSXnC3}
\end{figure}

Figure \ref{fig:3dHSXnC3} presents a graphical assessment of the \thwind{} optimization results and uses the color axis to highlight the regions where the strain is most severe. As can be seen the weak points are in the transition between bends, where some amount of \sbend{} and \stor{} is unavoidable. Comparing Fig. \ref{fig:tension2} and Fig. \ref{fig:3dHSXnC3}, these occur around bins 35, 190, and 330. The winding angle (pink vector in Fig. \ref{fig:3dHSXnC3}) changes angle by a significant amount at these points, yet there is still a finite bend radius.

\begin{figure}
\centering
\includegraphics[width=0.49\textwidth]{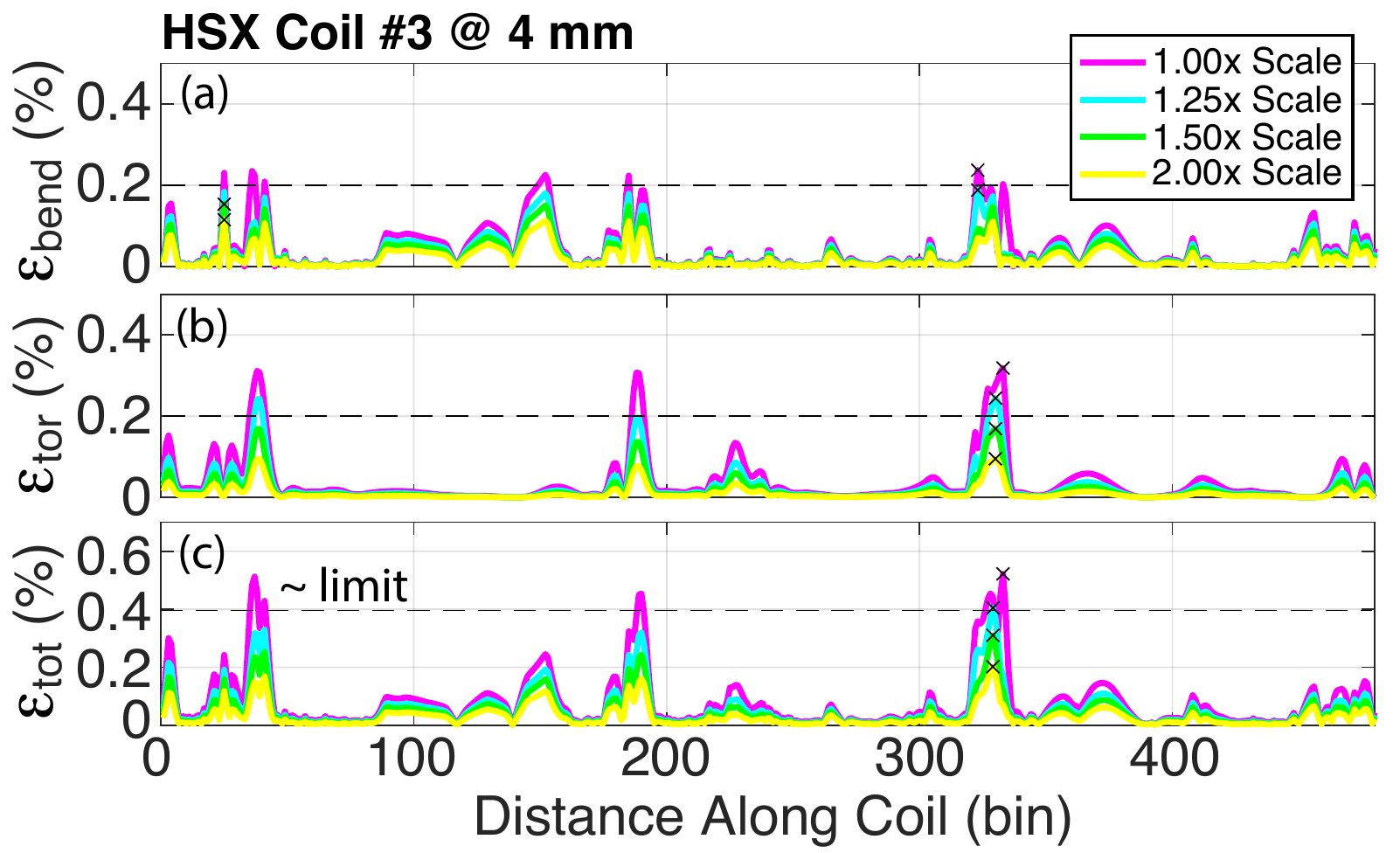}
\caption{Variation of the strain components as a function of \size{} for HSX coil \#3. For each \size{} the trajectory is optimized as in Fig. \ref{fig:tension2}. The maximum \stot{} (x-marks) naturally decreases with size.}
\label{fig:1dsizescale}
\end{figure}

The impact of \size{} on \sbend{}, \stor{}, and \stot{} is shown in Fig. \ref{fig:1dsizescale}, again using HSX coil \#3.  For each \size{}, optimization including possible manual intervention as described in Sec. \ref{sec:spline} has been undertaken. Despite optimization, it is found that the 1.0x size (as-built) coil exceeds the target strain of 0.4\%. As such, the as-built HSX is found to be too small to be compatible with the \uHTS{} strain limits as here assumed. Increasing the \size{} naturally reduces the strain, and already by 1.50x scale factor the strain is below the assumed limit.

\begin{figure}
\centering
\includegraphics[width=0.49\textwidth]{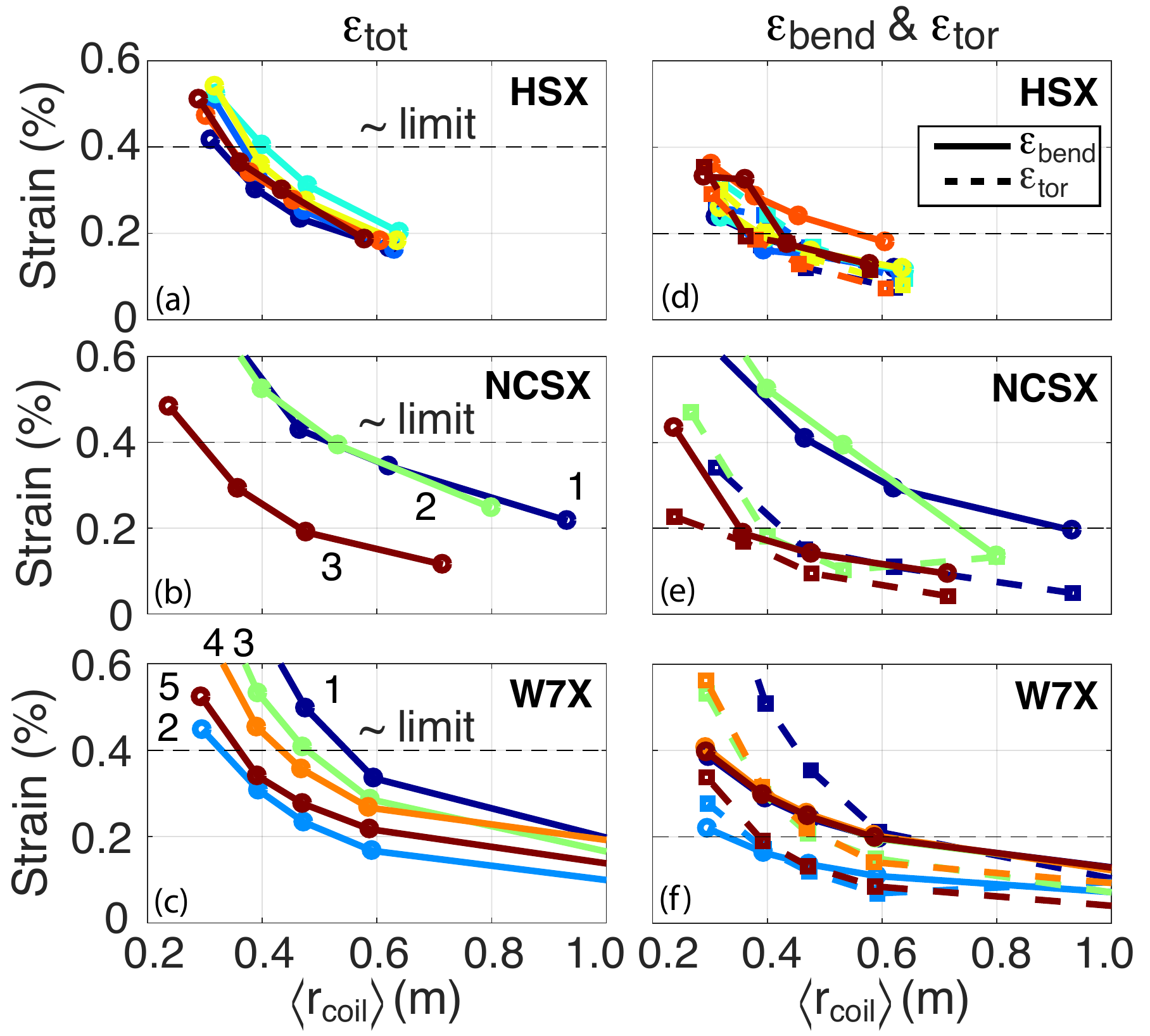}
\caption{(a-c) Peak total strain (\stot{}) and (d-f) peak bending (\sbend{}) and torsion (\stor{}) strain for the stellarator configurations as a function of average coil size (\ravg{}) for 4 mm wide tape. As \ravg{} decreases, the target \stot{} is exceeded, thus defining the minimum \ravg{}.}
\label{fig:strsum}
\end{figure}

Using the same methodology, strain assessment as a function of coil size (\ravg{}) was conducted for all coils of the HSX, W7-X, and NCSX stellarators. Results are presented in Fig. \ref{fig:strsum}. As coil size decreases, the target total strain is exceeded, thus defining the minimum buildable \ravg{} for these existing configurations. Generally, a minimum \ravg{} of 0.3 - 0.5 m is found, though variations between coils and configurations exist. 

Note that HSX coil \#3 is highlighted because it is most severely limited by strain, despite the fact that it is not the most non-planar. This implies the degree of non-planar complexity is not directly related to the strain limits encountered, and further suggests optimization of the coil trajectory itself has the potential to significantly improve compatibility with \uHTS{} coils. This is further discussed in Appendix \ref{sec:stellopt}.


\section{Combined Strain and Transverse Field Optimization}
\label{sec:both}

Optimizations considering cost functions involving both strain (\sbend{}, defined in Sec. \ref{sec:strlimit}) and transverse field (\Bperp{}, defined in Sec. \ref{sec:bperplimit}) using cost functions defined in Sec. \ref{sec:cost} are now presented. Coils optimized for both considerations must be larger than the minimum coil size (\ravg{}) shown in Fig. \ref{fig:strsum}, as headroom in strain is needed to trade-off against other factors (like \Bperp{}). Recall that to compute \Bperp{} the magnetic fields from all the coils comprising the configuration must be taken into account.

\begin{figure}
\centering
\includegraphics[width=0.49\textwidth]{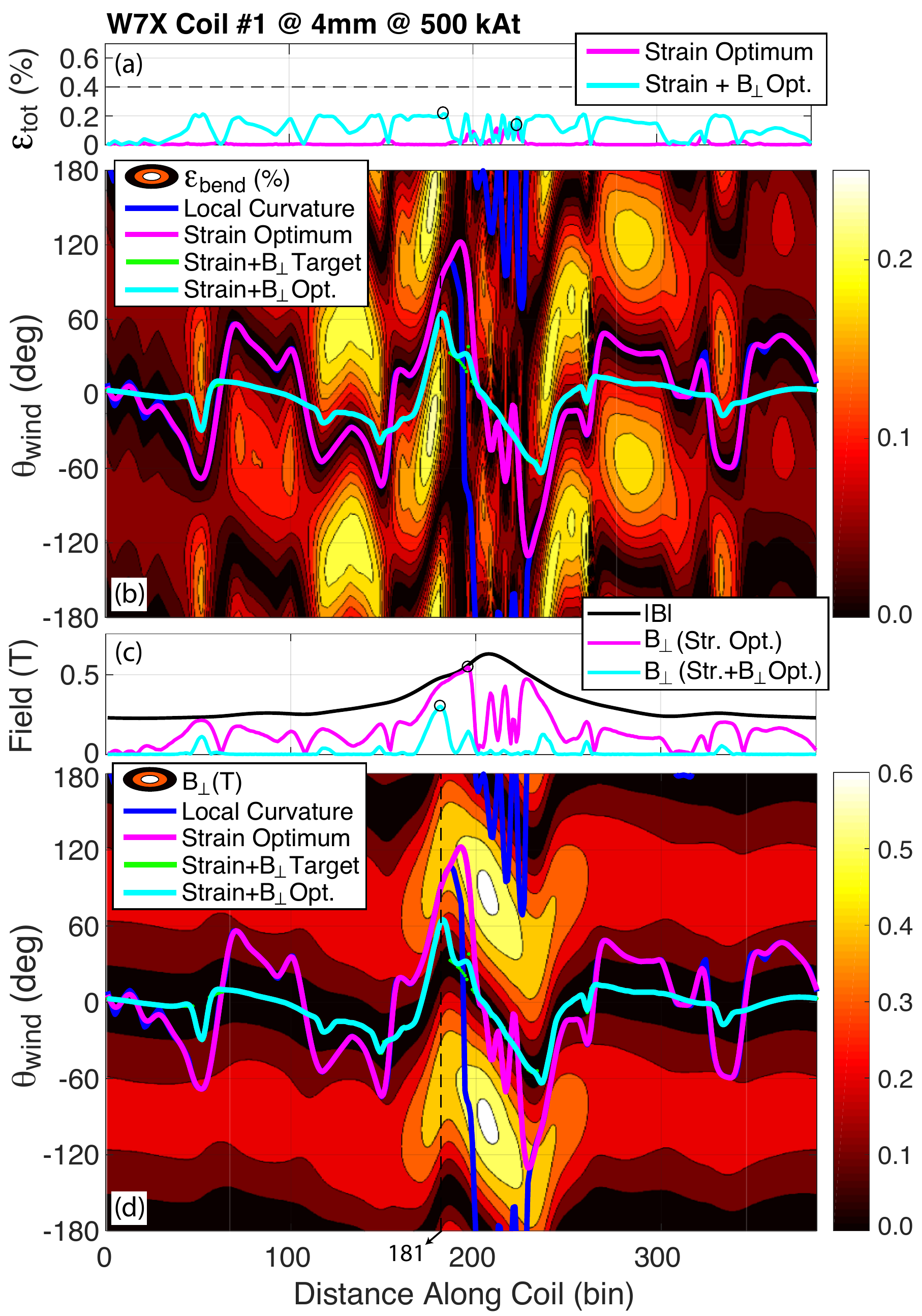}
\caption{Comparison of strain only (magenta) and combined strain+\Bperp{} (cyan) optimization for W7-X coil \# 1. Evaluations of (a) \stot{}, (b) \sbend{} vs. \thwind{}, (c) \Bperp{} and (d) \Bperp{} vs \thwind{}. Allowing finite \stot{} enables a significant reduction of \Bperp{} along the optimal \thwind{} trajectory.}
\label{fig:W7XnC1}
\end{figure}

Strain only (magenta) and combined strain + \Bperp{} (cyan) optimizations are demonstrated for coil \#1 of the W7-X stellarator configuration, with optimized trajectories shown in Fig. \ref{fig:W7XnC1}. For this larger coil the larger radii of curvatures yield \sbend{} contours that are significantly lower [Fig. \ref{fig:W7XnC1}(b)], enabling deviation of \thwind{} from the \sbend{} minimum. Contours of \Bperp{} [Fig. \ref{fig:W7XnC1}(c)] show a different dependency on \thwind{}. The combined optimization (cyan lines) follow the cost function target (green) very closely, essentially overlaying. For the combined optimization, \stot{} now takes a finite value for most of the trajectory [Fig. \ref{fig:W7XnC1}(a)], very close to the input \snot{} in Eq. \ref{eq:cost} value of 0.2 \%. The \Bperp{} value was also meaningfully reduced by this method, by nearly 50\%.

\begin{figure}
\centering
\includegraphics[width=0.49\textwidth]{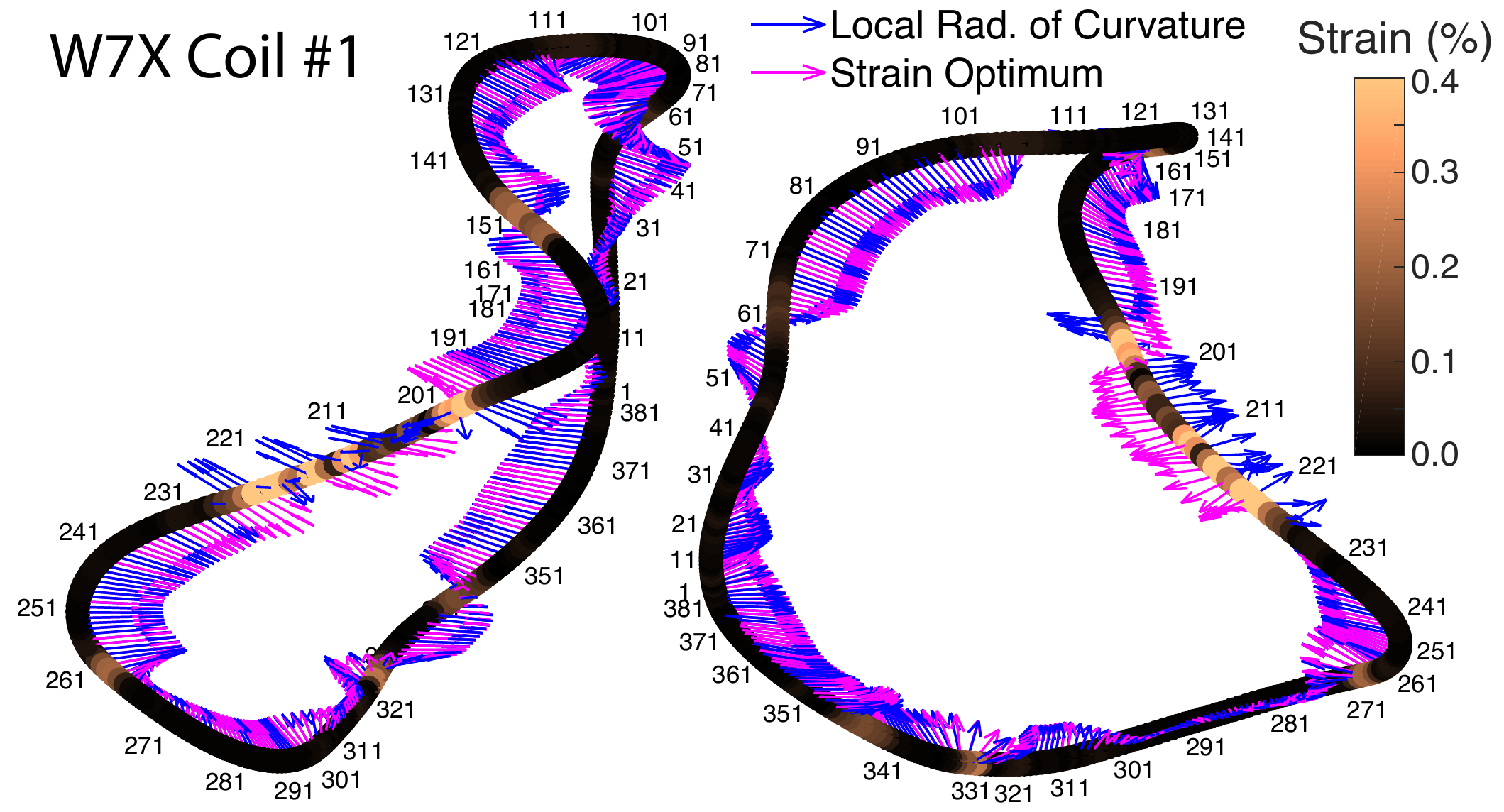}
\caption{Two viewing angles of W7-X coil \#1 including the local radius of curvature (blue vectors) and optimal \thwind{} trajectory (magenta vectors) for a strain-only optimization. Colors along the coil trajectory indicate relative \stot{}. Regions of high \stot{} are found at the straight section, indicating an artificial constraint is present.}
\label{fig:W7Xall_nC1_strOpt_3D}
\end{figure}

At this point it should be mentioned that some coils (such as the one highlighted in Fig. \ref{fig:W7XnC1}) contain apparent artefacts in the coil trajectory that inhibit compatibility with \uHTS{}.  This can be seen in the wiggles in the local curvature (blue line) in Fig. \ref{fig:W7XnC1}(b) around bin \#210. As \size{} is reduced, this feature imposes a high strain and limits the buildable size. As can be seen in Fig. \ref{fig:W7Xall_nC1_strOpt_3D}, this artefact occurs at the nominally straight section of the coil. While seemingly straight, these sectors are found to contain finite curvature (and finite \sbend{}) requiring significant torision (\stor{}) to mitigate. Improved coil trajectory definition should avoid these artefacts as will be described in Appendix \ref{sec:stellopt}.

\begin{figure}
\centering
\includegraphics[width=0.35\textwidth]{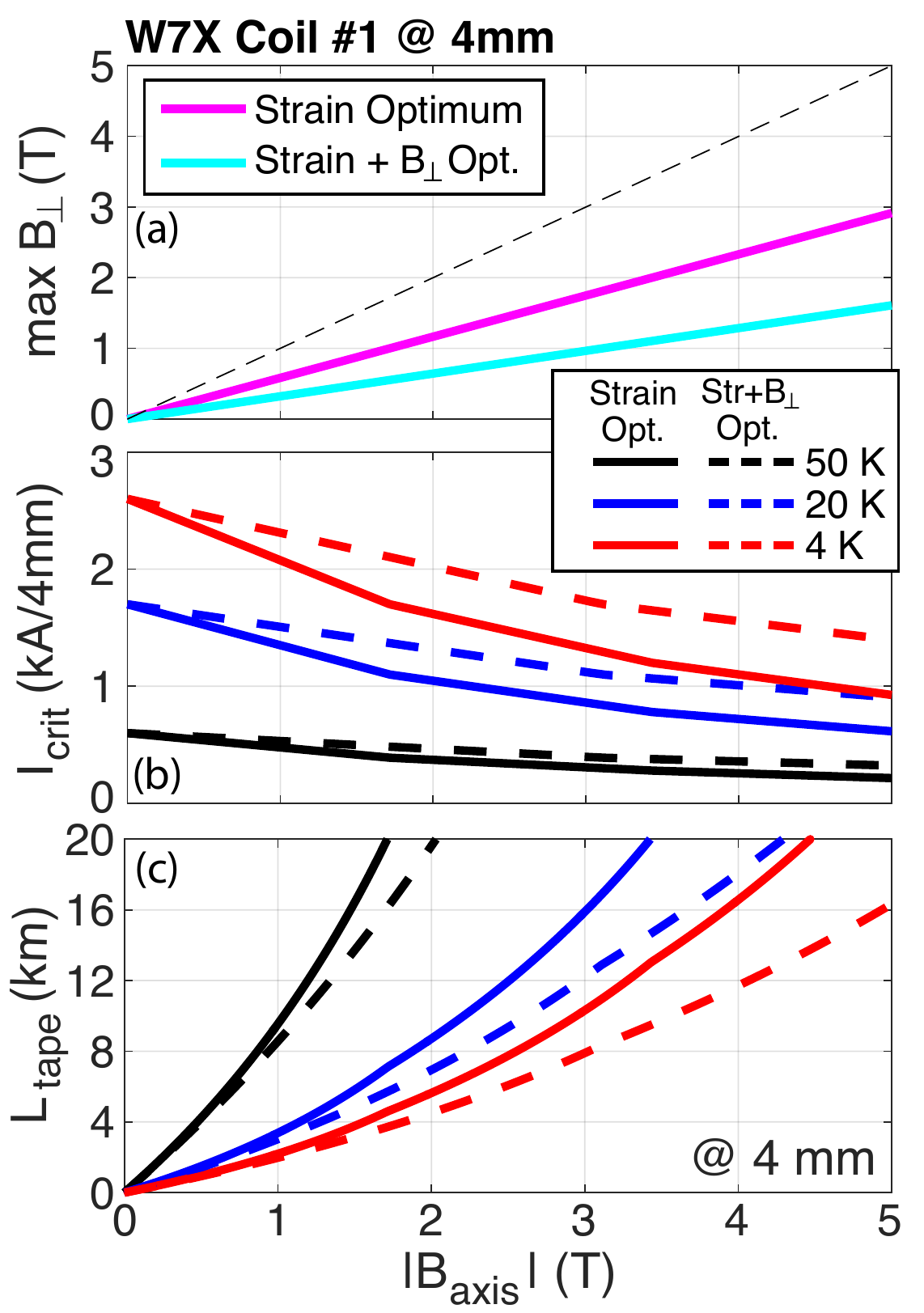}
\caption{(a) Transverse field (\Bperp{}), (b) critical current (\Icrit{}), and (c) required HTS tape length as a function of \Baxis{} for W7-X coil \#1.}
\label{fig:W7XnC1Bscale}
\end{figure}

Using publicly available data on the achievable \Icrit{} for a given HTS tape width at various \Bperp{} and operating temperature conditions \cite{Superpower}, the HTS tape length needed for a given \Baxis{} can be estimated. This is shown in Fig. \ref{fig:W7XnC1Bscale} for the same trajectories of Fig. \ref{fig:W7XnC1} using W7-X coil \#1. The reduction in \Bperp{} enables a meaningful increase in the achievable \Baxis{} for fixed tape width (\Ltape{}) or alternatively a reduction in \Ltape{} for a fixed \Baxis{}.

\begin{figure}
\centering
\includegraphics[width=0.49\textwidth]{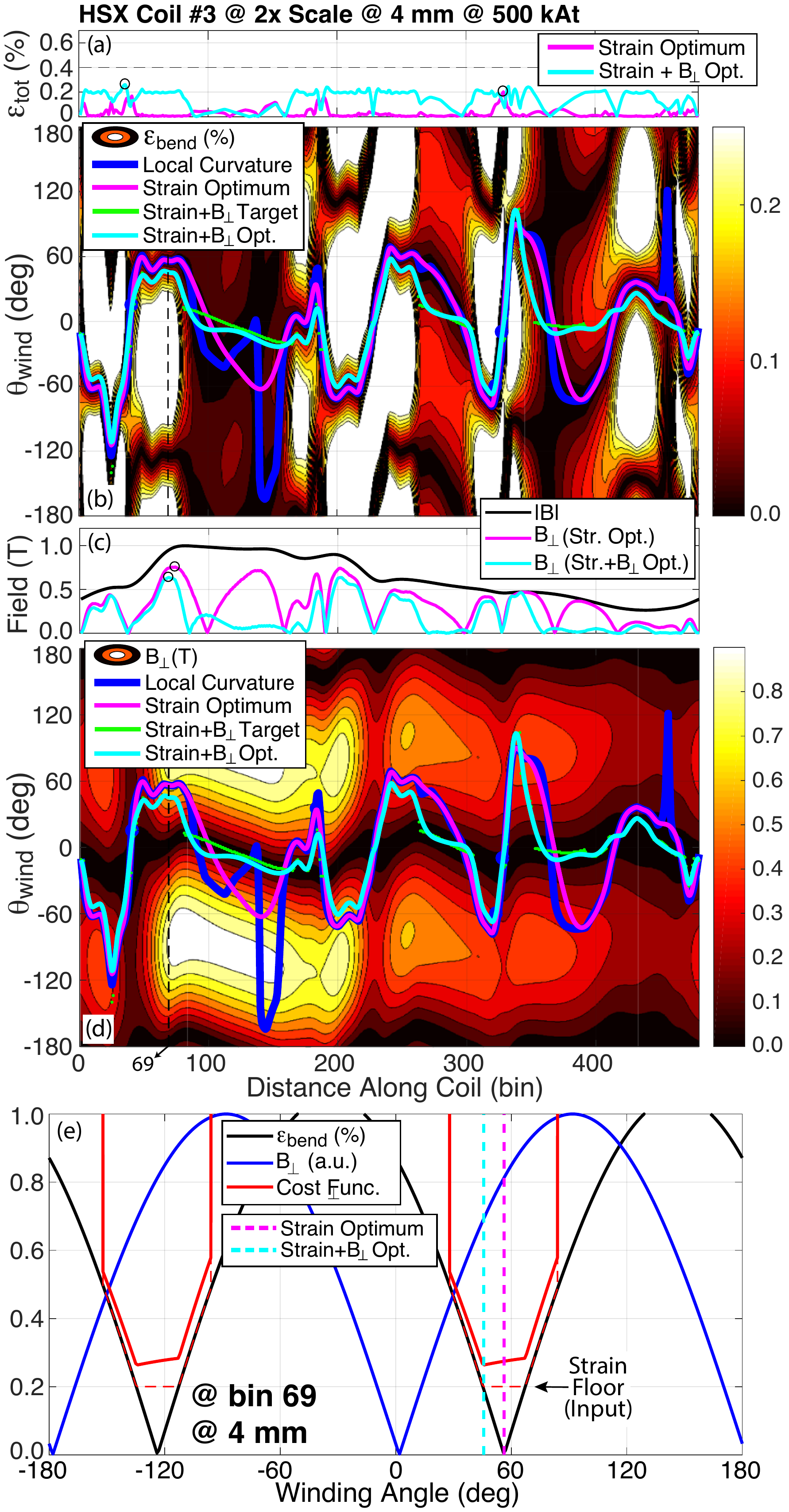}
\caption{Comparison of strain only (magenta) and combined strain+\Bperp{} (cyan) optimization for HSX coil \#3 at 2x \size{}. Evaluations of (a) \stot{}, (b) \sbend{} vs. \thwind{}, (c) \Bperp{} and (d) \Bperp{} vs \thwind{}. Allowing finite \stot{} is not found to improve this optimization by a significant degree, due to (e) a poor alignment of \Bperp{} and \sbend{} constraints around bin \#69.}
\label{fig:HSXnC3}
\end{figure}

\begin{figure}
\centering
\includegraphics[width=0.35\textwidth]{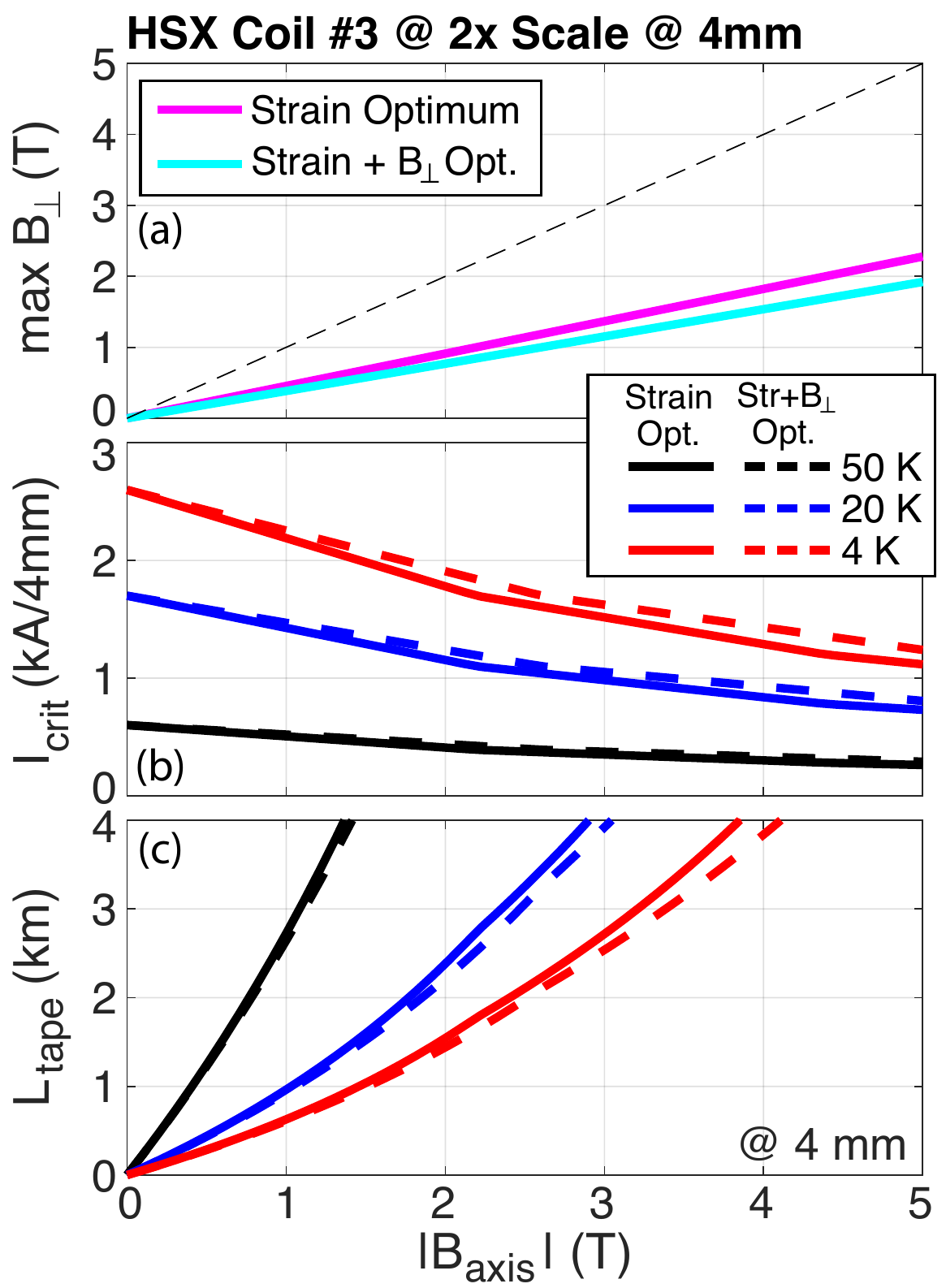}
\caption{(a) Transverse field (\Bperp{}), (b) critical current (\Icrit{}), and (c) required HTS tape length as a function of \Baxis{} for HSX coil \#3 at 2x \size{}. Combined strain + \Bperp{} optimization did not improve the required \Ltape{} in this case.}
\label{fig:HSXnC3Bcale}
\end{figure}

A second example is provided using the same HSX coil \#3 described in detail in Sec. \ref{sec:strain}. However, since the 1x \size{} was already above the target strain limit, a 2x \size{} is used. This provides the necessary headroom to in principle optimize against both strain and \Bperp{}. However, as shown in Fig. \ref{fig:HSXnC3}, allowing finite strain does not significantly improve optimization performance, with peak \Bperp{} is nearly unchanged. Looking in detail at the constrained region in Fig. \ref{fig:HSXnC3}(e), the \sbend{} is found to be below \snot{} only in a small region of \thwind{}. Within this allowable \thwind{} region, no significant \Bperp{} reduction can be achieved. Thus, this particular coil is resistant to further optimization.

Mapping of the \Icrit{} data to this coil as \Baxis{} is scaled is shown in Fig. \ref{fig:HSXnC3Bcale} for HSX coil \#3 at 2x \size{}. As \Bperp{} did not much change when included in the optimization, both strain only and combined yield similar results. Note this coil (even at 2x \size{}) requires significantly less \Ltape{} to achieve meaningful \Baxis{} due to its smaller size as compared to the W7-X coils.

\begin{figure}
\centering
\includegraphics[width=0.4\textwidth]{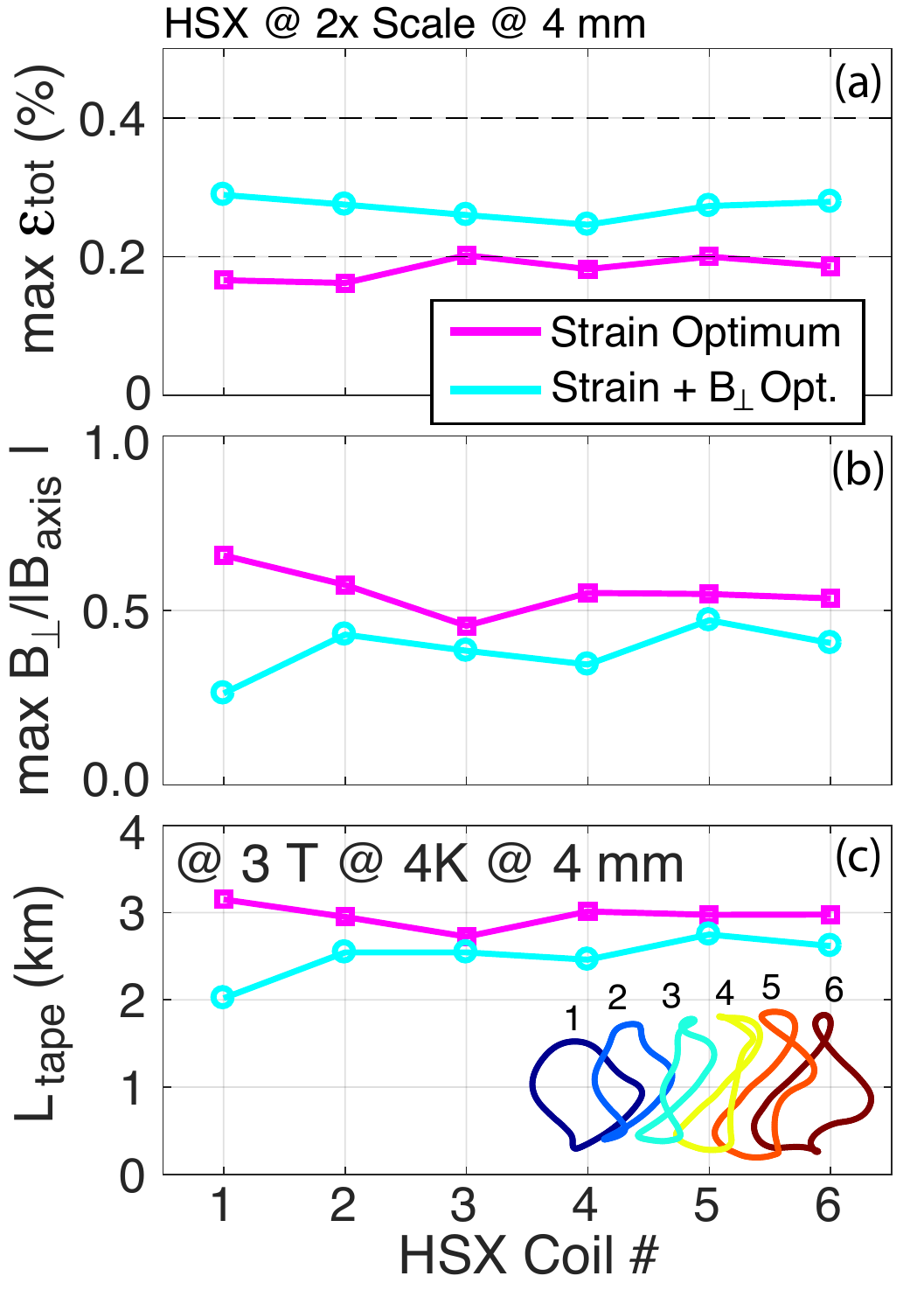}
\caption{Summary of performance improvement via combined strain + \Bperp{} optimization for all HSX coils at 2x \size{}. The most planar coils (\#1, 2) obtain a meaningful benefit while the more planar coils are fairly constrained and do not benefit as much. In both optimizations, \Ltape{} of a few km gives access to \Baxis{} of several T.}
\label{fig:BoptScaleSum}
\end{figure}

Combined optimization of all the coils in the HSX configuration at 2x \size{} is performed and results are given in Fig. \ref{fig:BoptScaleSum}. For many coils, the \Bperp{} component could be meaningfully reduced, especially for the least planar (coils \#1, 2). As discussed, the coil \#3 was barely affected, and the other most planar coils less-so. Nonetheless, at least in some instances the increased allowance for strain enables a significant reduction in the needed \Ltape{}. Final adjudication between all optimization constraints requires a target \Baxis{} as well as a notional budget, as increasing \Ltape{} implies an increased cost penalty.


\section{Discussion and Conclusions}
\label{sec:disc}

This work has presented the benefits and drawbacks of \uHTS{} magnet technology specifically for its application to non-planar coils. A novel winding angle optimization method is introduced to mitigate the drawbacks of increased \wrong{} bending strain (\sbend{}), torsional strain (\stor{}), and increased transverse field (\Bperp{}). By trading off the two strains against each other via an optimized \thwind{} trajectory, a minimum peak total strain and a reduced \Bperp{} can be obtained. This minimum peak total strain in turn enables assessment of the minimum buildable size for a given input non-planar coil geometry.  For well-known existing stellarator designs, the minimum size was found to be 0.3 - 0.5 m for 4 mm wide HTS tape. For coils larger than this minimum size, the total strain (\stot{}) can be traded off against \Bperp{} to reduce this component. This enables a reduction of the length of HTS tape required to achieve a given design magnetic field or equivalently an increase in the achievable magnetic field for constant HTS tape length.


\appendix
\section{Optimization of the Coil Trajectory for Stellarator Applications}
\label{sec:stellopt}

This work has focused on optimizing the winding angle optimization of a pre-defined coil to maximize compatibility with \uHTS{} magnet technology. Considerations for optimizing the coil trajectory itself now are briefly summarized. This discussion focuses on stellarator applications, as there are many possible degrees of freedom in the coil geometry of these concepts and significant coil optimization work already exists in this area \cite{Merkel1987,Pomphrey2001,Strickler2002,Zhu2018,Zhu2018a,Landreman2017,Paul2018}.

Stellarator coil optimizations are done with many constraints in mind, most having to do with plasma physics. Considering the constraints arising from the coil technology itself, two constraints are usually included: curvature and coil-coil distance. Interestingly, \uHTS{} magnets pose significantly different constraints than conventionally considered.

First, considering curvature, for a \uHTS{} magnet strain arising from regular curvature is negligible, and is also called `\easy{}' bending. The strain issues discussed in Sec. \ref{sec:strain} are important, and in particular the interplay of torsion and wrong-ways bending. If curvature is weakly penalized, yet torsion is not, very different stellarator coil shapes may arise from optimizations against these alternate criteria, particularly with more pronounced toroidal joggles.

Second, considering coil-coil spacing, the compactness (high current density) and mechanical strength (steel substrate and bobbin) inherent to the \uHTS{} magnet has the potential to support significantly reduced coil-coil spacing. The final spacing depends the amount of bobbin material required, which arises from the electromagnetic forces, which require definition of the target operating magnetic field. However, at least as compared to copper and LTS, significantly smaller coil-coil separations can be anticipated.

Finally, stellarator coils are usually parametrized via Fourier series. As for example in the W7-X coil of Fig. \ref{fig:W7Xall_nC1_strOpt_3D}, this gives rise to an artefact in the straight sections of the coil where residual undulations exist from incomplete cancellation of the Fourier series. These residual undulations severely compromise compatibility with \uHTS{}, despite their origin from a mostly-straight section of the coil. Using tensioned splines to parametrize the coil trajectory should remove this artificial limitation.

While outside the scope of this activity, stellarator coil optimizations using these alternate criteria are thus highlighted as fertile ground for future study.

\section*{Acknowledgments}
This work was supported by General Atomics Internal Funds. The author would like to thank A. Benson, B. Breneman, J. Leuer, J. Smith, Z.B. Piec, and L. Holland for useful discussions. The author also thanks S. Lazerson and A. Bader for the provision of the existing stellarator device coil geometry information.

\bibliographystyle{iopart-num}

\bibliography{library}


\end{document}